\documentstyle[12pt]{article}
\def\doit#1#2{\ifcase#1\or#2\fi}
\def\endtitle{\end{quotation}\newpage} 
\def\a{\alpha} \def\b{\beta}  \def\d{\delta}
\def\e{\epsilon}  \def\g{\gamma}
   
\def\l{\lambda} \def\m{\mu} \def\n{\nu} \def\o{\omega}
  \def\r{\rho} \def\s{\sigma}
\def\t{\tau}   
  \def\G{\Gamma}

\def\Tilde#1{{\widetilde{#1}}\hskip 0.015in}    
\def\Hat#1{\widehat{#1}}                        
\def\Bar#1{\overline{#1}}
\def\scst{\scriptstyle}
\let\du=\d 
\def\du#1#2{_{#1}{}^{#2}}   
\def\low#1{\hskip0.01in{\raise -2pt\hbox{${\hskip 1.0pt}\!_{#1}$}}}
\def\[{\lfloor{\hskip 0.35pt}\!\!\!\lceil\,}
\def\]{\,\rfloor{\hskip 0.35pt}\!\!\!\rceil}

\def\pl#1#2#3{Phys.~Lett.~{\bf {#1}B} (19{#2}) #3}
\def\pr#1#2#3{Phys.~Rev.~{\bf {#1}D} (19{#2}) #3}
\def\np#1#2#3{Nucl.~Phys.~{\bf B{#1}} (19{#2}) #3}
\def\cqg#1#2#3{Class.~and Quant.~Gr.~{\bf {#1}} (19{#2}) #3}
\def\mpl#1#2#3{Mod.~Phys.~Lett.~{\bf A{#1}} (19{#2}) #3}

\def\ul{\underline}
\def\un{\underline}
\def\tr{\,\,{\rm tr}}

\def\trz{\tr_z}
\def\fracm#1#2{\,\hbox{\large{${\frac{{#1}}{{#2}}}$}}\,}

\def\half{{\fracm12}}

\def\frac#1#2{{\textstyle{#1\over\vphantom2\smash{\raise -.20ex
\hbox{$\scriptstyle{#2}$}}}}} 
\def\ud#1#2{^{#1}{}_{#2}} 
\def\Lag{{\cal L}}                 

\def\-{{\hskip 1.5pt}\hbox{-}}
\def\nt{n\low T}
\def\nh{n\low H}
\def\nv{n\low V}
\def\vz{v^z} \def\tildevz{{\Tilde v}{}^z}

\let\la=\label
\def\nn{\nonumber}

\let\bm=\bibitem
\def\bd{\begin{document}}
\def\ed{\end{document}}
\def\be{\begin{equation}}
\def\ee{\end{equation}}
\def\ba{\begin{array}}
\def\ea{\end{array}}
\def\bea{\begin{eqnarray}}
\def\eea{\end{eqnarray}}
\newcommand{\eq}[1]{(\ref{#1})}
\newcommand{\hoch}[1]{$\, ^{#1}$}

\begin{document}
~\vspace{-1.0in}
\begin{flushright}
\hfill{UMDEPP 97-086}\\
CTP TAMU-14/97\\
\hfill{hep-th/9703075}\\
\hfill{\today}\\
\end{flushright}

\vspace{20pt}

\begin{center}

{\large\bf New Couplings of Six-Dimensional Supergravity} 

\baselineskip 10pt
\vskip 0.25in
Hitoshi ~N{\small ISHINO}\hoch{1}\\[.18in]
{\it Department of Physics} \\[.015in]
{\it University of Maryland} \\[.015in]
{\it College Park, 20742-4111, USA} \\[.25in]

and  \\[.25in]

Ergin ~S{\small EZGIN}\hoch{2}\\[.18in]
{\it Department of Physics and Astronomy} \\ [.015in]
{\it Texas A and M University}\\ [.015in]
{\it College Station, TX 77843-4242, USA} \\[.18in]

\end{center}

\vskip 0.5in

\begin{abstract}

\vskip 0.2in

We describe the couplings of six-dimensional supergravity, which
contain a self-dual tensor multiplet, to ~$\nt$~ anti-self-dual tensor
matter multiplets, ~$\nv$~ vector multiplets and ~$\nh$~
hypermultiplets. The scalar fields of the tensor multiplets form a coset
$~SO(\nt,1)/SO(\nt)$, while the scalars in the hypermultiplets form
quaternionic K\"ahler symmetric spaces, the generic example being
$~Sp(\nh,1)/Sp(\nh)\otimes Sp(1)$. The gauging of the compact subgroup
$~Sp(\nh) \times Sp(1)~$ is also described. These results generalize
previous ones in the literature on matter couplings of $~N=1$~
supergravity in six dimensions.

\end{abstract}

{\vfill\leftline{}\vfill
\vskip 10pt
\footnoterule
{\footnotesize
\hoch{1} Research supported in part by NSF Grant PHY-93-41926 and 
DOE Grant DE-FG02-94ER40854 \vskip -12pt}\vskip 10pt
{\footnotesize
\hoch{2} Research supported in part by NSF Grant PHY-9411543 
\vskip -12pt}}

\pagebreak
\setcounter{page}{1}

\centerline{\bf 1.~~Introduction}
\bigskip

Supersymmetric field theories in six dimensions (6D) are of considerable
interest for various reasons. Those which admit chiral supersymmetry,
namely the $(1,0)$ and $(2,0)$ supersymmetric cases, are especially
interesting, because when anomaly free, they may hint at significant
properties of various compactifications of a unifying theory in higher
dimensions, such as M-theory. 

While fairly general couplings of $(1,0)$ supergravity to matter were
described sometime ago by the authors \cite{ns}, recent investigations
of M-theory compactifications have unsurfaced interesting
generalizations of those couplings. For example, an anomaly free model
that contained nine anti-self-dual tensor multiplets, eight vector
multiplets and twenty hypermultiplets was found in \cite{sen} from
M-theory on $~(K_3\times S_1)/Z_2$. A low energy field theory for this
model does not exist in the literature at present. In this paper, we
shall close this gap and provide the most general couplings to date of
the $(1,0)$ supergravity in 6D.

As is well-known, when the numbers of self-dual and anti-self-dual tensor
multiplets are different, a manifestly Lorentz invariant lagrangian
formulation no longer exists. This is because the non-paired self-dual
or anti-self-dual components do not admit the usual kinetic term as the
square of the third-rank antisymmetric tensor field strength. In the
case of a single tensor matter multiplet coupling to the $(1,0)$
supergravity, a lagrangian formulation does exist, because of the paring
between the self-dual field strength of supergravity multiplet and the
anti-self-dual field strength of the matter tensor multiplet.

The $(1,0)$ supergravity will be alternatively referred to as $~N=1~$
supergravity. The first model of $~N=1$~ supergravity coupled to a
single tensor multiplet was given in \cite{ms}. A more complicated
system of $~N=1$~ supergravity coupled to a single tensor multiplet,
Yang-Mills vector multiplets, and hyper multiplets forming a quaternionic
K\"ahler manifolds was accomplished in terms of lagrangian formulation
in ref.~\cite{ns}.  (The model was called ~$N=2$~ supergravity in \cite{ns},
according to an alternative convention for counting the number of
supersymmetries.)  The gauging of the scalar manifold isometries, as well
as the $~Sp(1)~$ automorphism group was also given in \cite{ns}.

In a work by Romans \cite{romans}, the multiple tensor multiplets were
coupled to supergravity with no lagrangian formulation. It was found
that the tensor multiplets form a coset $~SO(\nt,1)/SO(\nt)$~ in order
for the couplings to supergravity to be consistent. Afterwards it was
found by Sagnotti \cite{sagnotti} that vector multiplets can be further
introduced to this system. In this case, an interesting Yang-Mills gauge
anomaly structure emerges already at the level of classical field
equations. This anomaly, and its relation to a supercurrent anomaly 
has been discussed in \cite{ferraraetal}.

Globally supersymmetric limits of the models mentioned above turn out to
be rather subtle. For example, the sigma model describing the
hypermultiplets is based on a hyper-K\"ahler manifold, rather than the
quaternionic K\"ahler manifold that arises in coupling to supergravity.
The coupling of a anti-self-dual tensor multiplet to Yang-Mills was
worked out in \cite{eee} by direct construction rather than a rigid
supersymmetry limit of the supergravity plus tensor multiplet system.
Such a limiting procedure is not known yet. Further peculiarities
arise in the tensor multiplet plus Yang-Mills system, for the
description of which, we refer the reader to \cite{eee}.

In this paper, we derive all the field equations $~N=1$~ supergravity
coupled to ~$\nt~$ tensor multiplets, ~$\nh$~ copies of hypermultiplets
and $~\nv$~ copies of vector multiplets. 
The ~$\nt~$ scalars of the tensor multiplet parametrize
the coset $~SO(\nt,1)/SO(\nt)$, and the $~4\nh~$ scalars of the
hypermultiplets parametrize the coset $~Sp(\nh,1)/Sp(\nh)\otimes Sp(1)$.
The choice of the latter coset is due to notational simplicity. Our
formulae can straightforwardly be adapted to more general quaternionic
symmetric spaces. In this paper, we also gauge the $~Sp(\nh)~$ subgroup
of the hyperscalar manifold isometry group $~Sp(\nh,1)~$ and the
automorphism group $~Sp(1)$. The supersymmetry transformations provided
here reduce to the full transformation rules of the single tensor
multiplet case, and in that sense we expect our result to be exact.
However, the fermionic field equations are given up to terms cubic in
fermions, and bosonic field equations up to fermionic bilinears. In the
interesting case of (anti) self-duality equations, the coefficients of
the fermionic bilinears are determined as well.

An interesting feature that emerges in the coupling of Yang-Mills to
multi-tensor multiplets is that the gauge kinetic term vanishes for
certain expectation value of the scalar fields \cite{sagnotti}. Here we
also find the correlated singularities in the full supersymmetry
transformation of the gravitino and the gaugino. It was proposed in
\cite{ts1} that these singularities signal a phase transition, and in
\cite{ts2,ts3} this was attributed to tensionless strings. The couplings
of the hypermatter constructed here do not exhibit singularities,
provided that the group $~Sp(\nh)\times Sp(1)$ is not gauged. The
gauging of this group, however, gives rise to singularities in a number
of hypermatter couplings at the point in the moduli space where the
previously known gauge coupling singularities occur. 

This paper is organized as follows. In the next section, we describe the
geometrical aspects of the scalar manifolds $~SO(\nt,1)/SO(\nt)$, and
$~Sp(\nh,1)/Sp(\nh)\otimes Sp(1)$, and set up the notation. In section
3, we give our results for all the field equations with supersymmetry
transformation rules, with their derivations based on mutual
consistency. In the same section, we also present the gauging of
$~Sp(\nh)\times Sp(1)$. The gauging of $~SO(\nt)~$ does not seem to be
possible for reasons that will be explained in section 3. In section 4,
the case of $~\nt=1$, namely a single tensor multiplet coupled to
supergravity, which admits a lagrangian formulation is presented.
Concluding remarks are given in section 5. The Appendix is devoted to
useful notations and conventions crucial for our computations.
\bigskip\bigskip

\centerline{\bf 2.~~Preliminaries}
\bigskip

We first fix the field contents of our total system. It consists of four
kinds multiplets: the multiplet of $~N=1$~ supergravity $~(e\du\m
m,\psi\du\m A)$, $~\nt$~ copies of anti-self-dual tensor multiplets plus
one self-dual tensor multiplet denoted collectively as $~(B_{\m\n}{}^I,
\chi^{A i}, \varphi^{\un\a})$, ~$\nv$~ copies of Yang-Mills vector
multiplets $~(A_\m, \l^A)$, and ~$\nh$~ copies of hypermultiplets
$~(\phi^{\a}, \psi^a)$. We use the world indices $~{\scst
\m,~\n,~\cdots~=~0,~1,~\cdots,~5}$~ and tangent space indices $~{\scst
m,~n,~\cdots~=~(0),~(1),~\cdots,~(5)}$. The indices $~{\scst
A,~B,~\cdots~=~1,~2}$~ label the fundamental representation of the
automorphism group $~Sp(1)$. The scalar fields $~\varphi^{\un\a}
\,{\scst (\un\a~=~\ul 1,~\cdots,~{\ul n}\low T)}~$ parametrize the coset
$~SO(\nt,1)/SO(\nt)$. The indices ~${\scst
I,~J,~\cdots~=~0,~1,~\cdots,~\nt}$~ label the fundamental representation
of $~SO(\nt,1)$, and the indices ~${\scst i,~j,~\cdots~
=~(1)~,~\cdots,~(\nt)}$~ label the fundamental representation of
$~SO(\nt)$. The hyperscalars $~\phi^\a\,{\scst(\a~=~1,~\cdots,~4\nh)}~$
parametrize the coset $~Sp(\nh,1)/Sp(\nh)\otimes Sp(1)$, and the indices
~${\scst a,~b,~\cdots~=~1, ~\cdots,~2\nh}$~ label the fundamental
representation of $~Sp(\nh)$.

The Yang-Mills multiplet fields are in the adjoint representation of a
product group $~G=G_1\times G_2\times G_3\times \cdots \times G_p$. Some
of these factors can be identified with any compact subgroups of the
isometry groups $~SO(\nt,1)~$ and $~Sp(\nh,1)$. In this paper we will
consider the case when $~G_1=Sp(\nh)~$ and ~$G_2=Sp(1)$. 

In describing the couplings of the tensor multiplet, it is useful to
introduce the coset representatives $~L_I~$ and $~L_I{}^i$, which together
form an $~(\nt+1)\times (\nt+1)~$ matrix which obeys the properties of an
$~SO(\nt,1)~$ group element \cite{gns}. Denoting the components of the 
inverse matrix by ~$L^I~$ and $~L_i{}^I$, they obey the relations
\be
L_I L^I = 1~~, ~~~~ L\du i I L_I = 0 ~~, ~~~~ L\du I i L^I = 0 ~~. \\
\ee
The $~SO(\nt,1)~$ invariant {\it constant} metric
\be
\eta\low{I J} \equiv - L_I L_J + L\du I i L_{J i} ~~, \la{eta}
\ee
can be used to raise and lower the ~$SO(\nt,1)~$ vector indices:
$~\eta\low{I J} L^J = - L_I,\ \eta\low{I J} L\du i J = L_{I i}$. 
Another useful tensor ~$~G_{I J}$~ is defined by
\be
G_{I J} \equiv  L_I L_J + L\du I i L_{J i}  ~~, \la{g} \\
\ee
with the important distinction with the sign in the first term compared
with \eq{eta}.  In contrast to the latter, $~G_{I J}$~ is {\it not} a constant
tensor, but it depends on the coordinates $~\varphi^{\un\a}$.

Composite $~SO(\nt)~$ connection $~A_{\un\a}{}^{i j}$, and coset vielbeins
$~V_{\un\a}{}^i~$ can be defined by
\be
\partial_{\un\a} L_I{}^i= -A_{\un \a}{}^i{}_j\ L_I{}^j + V_{\un \a}{}^i\ L_I\ .
\la{dl}\\
\ee
Thus we have the useful relations $~ D_{\un\a} L_I = \partial_{\un\a} L_I =
L\du I i V_{\un\a\,i}~, ~D_{\ul\a} L\du I i = L_I V\du{\un\a} i$,
where $~D_{\un \a}=\partial_{\un \a}+ A_{\un \a}$~ which yields the
commutator
\be
\[ D_{\un\a}, D_{\un\b} \] L\du I i =  \left( V\du{\un\a} j V\du{\un\b} i -
V\du{\un\b} j V\du{\un\a} i\right)\ L_{I j}\ .  \la{com} \\
\ee
The overall constant in the r.h.s., which is the square of the inverse
radius of the hyperboloid $~SO(\nt,1)/SO(\nt)$, is fixed to be +1 by the
$~H\chi\-$terms in the closure of two supersymmetries on
$~B_{\m\n}{}^I$. Furthermore, the curvature tensor can be read off from
\eq{com}, which shows that the manifold has a constant negative curvature.

The case $~\nt=1$~ is special in the sense that the original coset
$~SO(\nt,1)/SO(\nt)$~ is reduced to a semi-simple group manifold $~SO(1,1)$~
with the non-positive definite metric. Moreover, since there is a pair
of a self-dual and an anti-self-dual tensor multiplets forming a total
field strength free of (anti)self-dual condition, we can construct an
invariant lagrangian. We will discuss this particular case in section 4.

As for the coset $~Sp(\nh,1)/Sp(\nh)\otimes Sp(1)$, many of its properties
have been exhibited in \cite{ns}. It is useful to recall that given a
representative $~L$~ of this coset, the Maurer-Cartan form decomposes as
\be
L^{-1}\partial_\a L= A_\a{}^{a b} T_{a b} + A_\a{}^{A B} T_{A B} 
+ V_\a{}^{a A} T_{a A}\ ,
\ee
where $~T_{a b}$~ and $~T_{A B}$~ are generators of $~Sp(\nh)$~ and $~Sp(1)$,
$~T_{a A}$~ are the coset generators, $~A_\a{}^{a b}$~ and $~A_\a{}^{A B}$~
are the $~Sp(\nh)~$ and $~Sp(1)~$ composite connections, 
and $~V_\a{}^{a A}$~ are the coset vielbeins. It is convenient to 
define a triplet of complex structures $~J\du{\a\b}{A B}$~ 
as \footnote{A correction to a misprint in \cite{ns}: 
The r.h.s. of eq. (2.7) should be multiplied with ~$-2$.}  
\be
J_{\a\b}{}^{A B}= \left(V_{\a a}{}^A V_\b{}^{a B} 
+V_{\a a}{}^B V_\b{}^{a A} \right) \ ,
\ee
which obey the $~Sp(1)$~ algebra. The $~Sp(1)$~ curvature $~F_{\a\b}$~ is
related to the complex structures as 
\be
F_{\a\b}=2 J_{\a\b}\ . 
\ee
For completeness, we also record the relations obeyed by the vielbeins:
\be 
g\low{\a\b} V_{a A}{}^\a V_{b B}{}^\b = \e_{a b}\e\low{A B}\ , \quad\quad
V_{a A}{}^\a V^{a B\b} + {\scst \a \leftrightarrow \b} 
= g^{\a\b}\delta_A{}^B\ ,
\ee
\be
V_{a A}{}^\a V^{b A\b} + {\scst \a \leftrightarrow \b} 
= {1\over \nh} g^{\a\b}\delta_a{}^b\ ,
\ee
where $~\e^{a b}$~ and $~\e^{A B}$~ are the invariant tensors of $~Sp(\nh)$~ 
and $~Sp(1)$, respectively.

Let us next consider the local $~Sp(\nh)\times Sp(1)$~ gauge transformations
\be
\delta \phi^\a= \xi^\a{}_{a b}\Lambda^{a b}(x) 
+ \xi^\a{}_{A B}\Lambda^{A B}(x)\ ,
\ee
where $~\xi\du\a{a b}$~ and $~\xi\du\a{A B}$~ are Killing vectors in general,
but for the case at hand they take the simple form $~T^{a b}\phi_\a$~ and
$~T^{A B}\phi_\a$, respectively. The covariant derivative of the
hyperscalars are defined as 
\be
{\cal D}_\m \phi^\a= \partial_\m \phi^\a - g A_\mu{}^{a b}\xi\ud\a{a b}
-g' A_\mu{}^{A B} \xi^\a{}_{A B}\ ,
\ee
where $~g$~ and $~g'$~ are the gauge coupling constants for $~Sp(\nh)$~ and
$~Sp(1)$, respectively. 

The coupling of scalar field modifies the covariant derivatives of the
fermionic fields in such a way that the following replacements have to
be made
\bea
&& 
g A_\mu{}^{a b}  \rightarrow g A_\mu{}^{a b}+({\cal D}_\m\phi^\a) A_\a{}^{a b}
\quad\quad {\rm (except\ in }\ D_\mu \lambda)~~ , \nn\\
&& 
g' A_\mu{}^{A B}  \rightarrow g' A_\mu{}^{A B}
+({\cal D}_\m\phi^\a) A_\a{}^{A B}~~.
\la{rs}
\eea
The reason for the exception made for the covariant derivative of $~\l$~
is a technical one, and it is explained in \cite{ns}. One
consequence of the above replacements is that the occurrence of Yang-Mills
field strength dependence terms in the following commutator
\be
\[{\cal D}_\mu, {\cal D}_\nu\]\epsilon^A = \fracm 1 4 R_{\mu\nu}{}^{mn}
\gamma_{mn} \epsilon^A
+({\cal D}_\mu\phi^\a) ({\cal D}_\nu\phi^\b) \ F_{\a\b}{}^{A B} \epsilon\low B
-\tr_z F_{\mu\nu} \ C^{A B}\epsilon\low B\ , \la{gcom}\\
\ee
where the triplet of functions $~C^{A B}$~ lie in the $~Sp(\nh)\times Sp(1)$~
algebra, and is given by
\be
C^{A B}= g A_\a{}^{A B}\xi^{\a c d}\ T_{c d}
+ g'\left( A_\a{}^{A B}\xi^{\a C D}\ T_{C D}-
T^{A B}\right)\ . \la{c}
\ee
This function arises in the field equations as well as the supersymmetry
transformation rules, as we shall see in the next section. As discussed
in detail in \cite{ns}, this function satisfies 
\be
{\cal D}_\m C^{A B} = \left( {\cal D}_\m \phi^\a \right) 
\left( {\cal D}_\a C^{A B} \right)
\ , \la{cr} \ee
and one can derive \cite{ns}
\be
{\cal D}_\a C^{A B} = 2 J_{\a\b}{}^{A B} \xi^\b\ ,  \la{dc}
\ee
where
\be
\xi^\a \equiv g A_\a{}^{A B}\xi^{\a c d}\ T_{c d} 
+ g' A_\a{}^{A B}\xi^{\a C D}\ T_{C D}\ .
\ee

One of the peculiar features of the vector couplings to the tensor
multiplet is the necessity of a constant matrix $~C^{I z}$, where the
index ~$^z$~ distinguishes the various factor groups in the total
Yang-Mills gauge group, while $~{\scst I ~=~ 1,~2,~\cdots,~\nt}$~ is for
the local coordinate index for the coset $~SO(\nt,1)/ SO(\nt)$~
\cite{sagnotti}. The constant coefficients $~C^{I z}$~ have been related to
certain $~S$-matrix elements in the conformal field theory an open
superstring \cite{sagnotti}. It is convenient to define the following
quantities which arise frequently in our calculations and results: 
\be
C^z \equiv C^{I z}L_I\ , \quad\quad\quad  C^{i z}
\equiv C^{I z}L_I{}^i\ , \la{d1}
\ee
\be
H_{\m\n\r} \equiv  H_{\m\n\r}{}^I L_I \ , \quad\quad\quad
H_{\m\n\r}{}^i\equiv H_{\m\n\r}{}^I L_I{}^i \ . \la{d2} 
\ee

\bigskip\bigskip

\centerline{\bf 3.~~The Field Equations and Supersymmetry Transformations}
\bigskip

Our strategy is to start with a general ans\"atze for the supersymmetry
transformations and field equations with unknown coefficients. We then
determine all the coefficients by the closure of supersymmetry
transformations, modulo filed equations when necessary, and the
requirement for the field equations to transform into each other. Although
we will give the bosonic field equations up to fermionic bilinears and
fermionic field equations up to cubic in fermion terms, it turns out
that we can still fix the full transformation rules, as well as the full
(anti)self-duality equations. We shall first give the results, and then
explain the derivations. Several formulae that are useful in these
derivations are provided in the Appendix.

The fermionic field equations are \footnote{A correction to a misprint
in \cite{ns}: The gaugino field equation in eq. (4.3) should have the
additional term: ${\sqrt2} e^{-\varphi/{\sqrt2}} \psi^a V_{\a a A}
\Tilde \xi^{\a\Hat I}$, in notation of \cite{ns}.}
\bea
&&
\g^{\m\n\r} {\cal D}_\n \psi_\r{}^A +\fracm 1 2 H^{\m\n\r} \g_\n \psi_\r{}^A
-\fracm 1 2 \g^\n\g^\m \chi^{Ai} V_{\un\a}{}^i \partial_\n\varphi^{\un\a}
 -2 \g^\n \g^\m \psi_a  V\du\a{a A} {\cal D}_\n \phi^\a \nn\\[0.01in]
&& 
~~~~~ + \fracm 1 2 C^z \tr_z \left( \g^{\r\s} \g^\m \l^A  F_{\r\s}\right)
+ \fracm 1 4 H^{\m\n\r}{}_i \g_{\n\r} \chi^{Ai}
- \tr_z \left( \g^\m C^{A B}\l_B\right)= 0 ~~, \la{f1}\\[0.15in]
&&
\g^\m  {\cal D}_\m \chi^{Ai} - \fracm1{24} \g^{\m\n\r}
\chi^{Ai}  H_{\m\n\r} -\half C^{i z} \tr_z\left(
\g^{\m\n} \l^A F_{\m\n} \right)  
+\fracm 1 4 H^{\m\n\rho i} \g_{\m\n}\psi_\rho{}^A \\[0.02in]
&&
~~~~~ -\fracm 1 2 \g^\m\g^\n\psi_\m  V_{\un\a}{}^i \partial_\n \varphi^{\un\a}
- C_z^{-1} C^{i z} 
\tr_z \left(C^{A B}\l_B\right)=0 ~~ , \la{f2}\\[0.15in]
&&
\g^\m {\cal D}_\m \psi^a + \fracm1{24} \g^{\m\n\r} \psi^a H_{\m\n\r}
-\g^\m \g^\n \psi_{\m A} V_\a{}^{a A} {\cal D}_\n \phi^\a 
 -2\l_A{}\,\xi^\a\, V_\a{}^{a A} = 0 ~~, \la{f3}\\[0.15in]
&&  
C^z \g^\m {\cal D}_\m \l_A + \fracm 1 4  C^{i z}
\g^{\m\n} \chi_{Ai} F_{\m\n}
+ \fracm1{24} C^{i z} \g^{\m\n\r} \l_A H_{\m\n\r i}
+ \half C^{i z} \g^\m \l_A V_{\un\a i} \partial_\m\varphi^{\un\a}\nn\\[0.02in]
&&
~~~~~ +\fracm 1 4 \g^\m \g^{\n\rho} \psi_{\m A} F_{\n\rho}
+\half C_{A B}\g^\m\psi_\m{}^B -\half C_z^{-1} C^{i z} C_{A B} \chi_i{}^B 
-2 \psi^a V_{a A}{}^\a\,\xi_\a = 0 ~~, \la{f4} 
\eea
where $~T_z$~ are the generators of the algebra in the adjoint
representation of the gauge group labelled by $~_z~$ and the summation
over ~${\scst z}$~ is always understood. All the terms which involve the
gravitino field $~\psi_\mu{}^A$~ result from supercovariantizations.

The bosonic field equations, up to fermionic bilinears, are
\bea
&& 
H_{\m\n\r}^+ =0 ~~, \la{b1}\\[0.15in]
&&
H_{\m\n\r}^-{}^i= 0 ~~, \la{b2}\\[0.15in]
&& 
R_{\m\n} =  \fracm1 4 G_{I J} H_{\m\r\s} {}^I H\du\n{\r\s\, J}
+g_{\un\a\un\b} \partial_\m\varphi^{\un\a} \partial_\n\varphi^{\un\b} 
+ 4 g\low{\a\b} \left( {\cal D}_\m \phi^\a\right) 
\left({\cal D}_\n \phi^\b \right)  \nn\\[0.02in]
&& 
~~~~~ ~~~\, + 2 C^z \tr_z \left(F\du\m\r F_{\n\r} - \fracm 1 8 g_{\m\n} 
F_{\r\s}{} ^ 2
\right) + \fracm 1 4 g_{\m\n} C_z^{-1}\tr_z \left( C^{A B}C_{A B}\right) ~~,
\la{b3}\\[0.15in]
&& 
e^{-1}D_\m \left( e g^{\m\n} \partial_\n\varphi^{\un\a}\right)
+\Gamma_{\un\b\un\g}{}^{\un\a}
\partial_\m\varphi^{\un\b} \partial^\m\varphi^{\un\g}
 -\half V_i{}^{\un\a} C^{i z} \tr_z F_{\m\n} F^{\m\n}\nn\\[0.02in]
&& 
~~~~~ -\fracm 1{6} V_i{}^{\un\a} H_{\m\n\r} ^i H^{\m\n\r}
- V\du i{\un\a} C_z^{-2}C^{i z} \tr_z \left( C^{A B} C_{A B} \right) = 0 ~~,
\la{b4}\\[0.15in]
&&
e^{-1}{\cal D}_\m\left( e g^{\m\n} {\cal D}_\n \phi^\a \right)
+ \Gamma_{\b\g}{}^\a~\left({\cal D}_\m\phi^\b\right) 
\left({\cal D}^\m\phi^\g \right) 
- C_z^{-1} J^\a{}_\b{}_{A B} \tr_z \left(C^{A B} \xi^\b\right)= 0~~,
\la{b5}\\[0.15in]
&&
{\cal D}_\n \left(e C^z F^{\m\n}\right)
+ \fracm 1 2 e \left( C^z H^{\m\r\s}_-+ C^{i z} H^{\m\r\s}_+{}_i \right) 
F_{\r\s} -2  e \xi_\a {\cal D}_\m\phi^\a  = 0 \ .\la{b6}
\eea

From experience with the $~\nt=1$~ case, we expect that all the higher order
fermion terms, except those which involve only matter fermions, can be
determined by supercovariantization of the field strengths and covariant
derivatives. In the interesting case of (anti) self-duality equations \eq{b1}
and \eq{b2}, we can actually determine them exactly. We find
\bea
\Hat H_{\m\n\r}^+ &=& \fracm14 \left( {\Bar\chi}{}^i 
\g_{\m\n\r} \chi\low i\right)
+ \half \left(\Bar\psi{}^a \g_{\m\n\r}\psi_a \right) ~~, \la{dual1}\\[0.15in]
\Hat H_{\m\n\r}^-{}^i &=& -\half C^{i z} \tr_z \left(\Bar\l\g_{\m\n\r}\l\right)
~~, \la{dual2}
\eea
where the (anti) self-dual field strengths are defined as the
suitable projections of the supercovariantized field strength
\bea
&& \Hat H_{\m\n\r}{}^I \equiv 3\partial_{\[\m} B_{\n\r\]} {}^I
+ 3 C^{I z} \tr_z \left( F_{\[\m\n} A_{\r\]}
-\fracm 1 3 \[ A_{\[\m} , A_{\n} \] A_{\r\]} \right)   \nn\\
&& ~~~~~ ~~~~~ ~~
+ 3 \left( \Bar\psi_{\[\m} \g_\n \psi_{\r\]} \right) L^I
- 3 \left(
\Bar\psi_{\[\m} \g_{\n\r\]} \chi^i \right) L\du i I ~~.\la{sc1}
\eea
The gauge coupling constant is suppressed in the definition of the
Chern-Simons form, for simplicity in notation. Despite the fact that the
field equations given above are up to higher order fermionic terms,
interestingly enough one can determine the full supersymmetry
transformation rules. Using the the previous works
\cite{ns,romans,sagnotti} as a guideline, we find the following
generalized result:
\bea
&& 
\d e\du\m m = (\Bar\e\g^m\psi_\m)~~, \la{t1}\\[0.1in]
&& 
\d \psi_\m = {\cal D}_\m(\Hat \omega) \e + \fracm1{48}
\g^{\r\s\t} \g_\m  \e \Hat H_{\r\s\t} 
- (\d \phi^\a ) (A_\a \psi_\m) \nn\\[0.02in]
&& 
~~~~~ ~~~- \fracm 1{16} \g_\m \chi^i (\Bar\e \chi_i) - \fracm 3{16}
\g_\n \chi^i (\Bar\e \g_{\m\n} \chi_i)
+ \fracm 1{32} \g_{\m\n\r} \chi^i (\Bar\e \g^{\n\r} \chi_i)\nn\\[0.02in]
&& 
~~~~~ ~~~- \fracm 9 8 C^z \tr_z \left[\, \l (\Bar\e\g_\m\l) \, \right ]
+ \fracm 1 8 C^z \tr_z \left[\, \g_{\m\n} \l 
(\Bar\e\g^\n \l)\, \right] \nn\\[0.02in]
&& 
~~~~~ ~~~ - \fracm 1{16}  C^z \tr_z \left[\, \g_{\r\s} \l
(\Bar\e \g\du\m{\r\s} \l) \, \right] + \fracm 1 {16} \g^{\n\r}
\e \, (\Bar\psi{}^a \g_{\m\n\r} \psi_a) ~~, \la{t2}\\[0.15in]
&& 
\d B_{\m\n}{}^I = 2 C^{I z} \tr_z \left( A_{\[\m}
\d A_{\n\]} \right) -2 \left(\Bar\e \g_{\[\m}
\psi_{\n\]} \right) L^I + \left(\Bar\e\g_{\m\n}
\chi^i \right) L\du i I ~~, \la{t3}\\[0.1in]
&&
\d \varphi^{\un\a}=(\Bar\e\chi^i) \, V_i{}^{\un\a} ~~,\la{t4}\\[0.1in]
&& 
\d \chi^i = \half \g^\m \e \Hat D_\m\varphi^{\un\a} V_{\un\a}{}^i
- \fracm 1{24} \g^{\m\n\r} \e
\Hat H_{\m\n\r}{}^i- (\d \phi^\a) (A_\a\chi^i)
- (\d\varphi^{\un \a}) A\du{\ul \a}{i j} \chi_j  \nn\\[0.02in]
&& 
~~~~~ ~~~ + \half C^{i z} \tr_z\left[\, \g^\m\l
\left( \Bar\e\g_\m \l \right) \, \right]  {~~,~~~~~} \la{t5}\\[0.15in]
&& 
\d A_\m{}=(\Bar\e \g_\m \l)~~,\la{t6}\\[0.1in]
&&
\d \l_A = - \fracm 1 4 \g^{\m\n} \e\low A \Hat F_{\m\n}
 - (\d \phi^\a ) (A_\a \l_A) - C_z^{-1} C^{i z}
\left(\Bar\chi_{i(A}\l_{B)} \right) \e^B -\half C_z^{-1} C_{A B} \e^B {~~,~~~}
\la{t7}\\[0.15in]
&& 
\d\phi^\a = V\du{a A} \a \left(\Bar\e^A \psi^a\right) ~~, \la{t8}\\[0.1in]
&& 
\d\psi^a = \g^\m \e\low A \Hat {\cal D}_\m \phi^\a\ V\du\a{a A} - (\d\phi^\a)
(A_\a \psi)^a ~~. \la{t9}
\eea
Here all the {\it hatted} field strengths and covariant derivatives are
supercovariantizations of the {\it non-hatted} ones. Recalling that
supercovariantization is achieved by replacing the
parameter $~\e^A$~ in the supersymmetry transformation by
$~(-\psi_\m{}^A)$, in addition to \eq{sc1}, one finds
\bea
&&
\Hat F_{\m\n}  \equiv F_{\m\n} 
- 2 \left( \Bar\psi{}_{\[\m} \g_{\n\]} \l  \right) ~~, \\
&& 
\Hat D_\m \varphi^{\un\a} \equiv \partial_\m \varphi^{\un\a}
-V\du i{\un\a}\left(\Bar\psi_\m \chi^i \right) ~~, \nn\\
&& 
{\Hat{\cal D}}_\m \phi^\a \equiv {\cal D}_\m \phi^\a
- V\du{a A} \a \left(\Bar\psi\du\m A \psi^a \right) ~~, \la{sc2}
\eea
The supercovariant derivative occurring in \eq{t2} is defined by
\be
{\cal D}_\m(\Hat \omega) \epsilon^A \equiv  \left[ \, \partial_\m 
\varepsilon^{A B} +\fracm 1 4 \Hat\omega_\m{}^{rs} \g_{r s}\varepsilon^{A B} 
+ g' A_\m{}^{A B} +({\cal D}_\m\phi^\a)A_\a{}^{A B}\,\right]\,\epsilon_B\ , 
\la{cd}
\ee
where the supercovariantized spin connection is given by
\be
\Hat\omega_{m r s} = \frac 1 2 (\Hat C_{m r s} - \Hat C_{m s r} 
+ \Hat C_{s r m})\ , 
\ee
and $~\Hat C$~ is supercovariantized Ricci's rotation coefficient:
\be
 \Hat C_{\m\n m} \equiv \partial_\m e_{\n m} - \partial_\n e_{\m m}
- (\bar\psi_\m \g_m \psi_\n) ~~.  
\ee

Note that the gravitational constant has always been
suppressed. It can easily be reintroduced by assigning mass dimension
$~1$~ to bosons, $~3/2$~ to fermions and $~-1/2$~ to $~\e$. Since the
supersymmetry transformations determined here reduce to the {\it full}
supersymmetry transformations given in \cite{ns} for the ~$\nt=1$~ case,
we conjecture that they are the full supersymmetry transformations for
all $~\nt$.

The supersymmetry transformation rules presented above form a closed
algebra with the composite parameters $~l_{m n}$~ for the Lorentz
transformation, $~\e_3$~ for the supersymmetry transformation, and
$~\xi^\m$~ for the general coordinate transformation given by
\bea
&&
\[ \d(\e_1), \d(\e_2) \] e_{\m m}  = \xi^\n \partial_\n e_{\m m} +
(\partial_\m \xi^\n) e_{\n m} + (\Bar\e{\low 3} \g_m \psi_\m)
+ l\du m n e_{\m n} ~~, \nn\\[0.02in]
&& 
\xi^\m \equiv \left( \Bar\e_2\g^\m\e_1 \right)~~, \nn\\
&&
\e_3^A  \equiv - \xi^\m \psi_\m + \left[\,  V\du{b B}\a (\Bar\e_2^B
\psi^b ) (A_\a \e_1)^A - {\scst  (1 \leftrightarrow 2) } \, \right] ~~,\nn\\
&&
l_{m n} \equiv \xi^\m \Hat\omega_{\m m n} + \bigg[ \fracm 1{24}
\left(\Bar\e_2 \g_{\[ m} \g^{\r\s\t} \g_{n\]}\e_1 \right) H_{\r\s\t}^-{}^I
L_I  - \fracm 1 8 \xi^\m \left( \Bar\psi{}^a \g_{\m m n} \psi_a \right)
\nn\\
&& 
~~~~~ ~~~~~ ~~~~~ ~~~~~ ~~~~~ - \fracm1 4 \left( \Bar\e_2 \g_{m n} 
\chi^i \right) \left( \Bar\e_1 \chi_i \right)  - \fracm 1 8 \left(
\Bar\e_2 \g\du{\[ m} \r \chi^i  \right) \left( \Bar\e_1 \g_{n\]\r} \chi_i
\right) \nn\\
&&
~~~~~ ~~~~~ ~~~~~ ~~~~~ ~~~~~  - C^z
\tr_z\left( \Bar\e_2 \g_{\[ m}\l \right) \left(\Bar\e_1\g_{n\]}\l\right)\bigg]
- {\scst (1\leftrightarrow 2) } ~~.\la{closure}
\eea

We now outline the steps we have followed to derive the equations of
motion and supersymmetry transformations.

(1) We first parametrize the transformation rules and the
(anti) self-duality equations, as dictated by the symmetries
of the theory, the existing partial results for the multi-tensor
multiplet couplings \cite{romans}, and the full results for the $~\nt=1$~
case \cite{ns}. At first, we assume that all matter fields are inert under the
{\it local} Yang-Mills gauge transformations, as well as the {\it local}
$~Sp(1)$~ gauge transformations.  It is convenient to perform the gauging 
process, after the ungauged results are obtained.

Note that factors of $~C_z^{-1}$~ occur in a number of places in the
equations of motion and the transformations rules. While these factors
may seem unusual, it is easy to understand their origin, which has to do
with the fact that we have parametrized the field equations for the
gauge fermion and the Yang-Mills field (excluding the hypermatter
contributions which are presented here) in such a way that they agree
with those of ref.~\cite{sagnotti}. For example, once the $~C^z$~ factor
is introduced in \eq{f4}, it is clear that the closure of the
supersymmetry transformations \eq{t7} must include the $~C^{-1}C^i
{\Bar\chi}\low i\l\-$term in order to produce the $~C^i\l H\-$term in
the $~\chi^i\-$field equation, upon the variation of $~\chi^i$. In
comparing this term with the $~\nt=1$~ case \cite{ns}, it is useful to note 
the identity \eq{lc} provided in the Appendix.

(2) Next, we require the closure of the supersymmetry transformations on
the {\it bosons}. Normally, closure on the bosons does not require any
field equations. However, as it has been known for some time
\cite{schwarz}, in the case of self-duality conditions which serve as
equations of motion, the closure of supersymmetry algebra on the bosons
does require the (anti) self-duality equations. Completing the closure
calculation on the bosons, we are able to fix all the supersymmetry
transformations, including the $~(\hbox{fermion})^2\-$terms, as well as
the (anti) self-duality equations \eq{dual1} and \eq{dual2}. In this context,
\eq{eb} given in the Appendix is useful in establishing the closure on
$~e_\m{}^m$~ and $~B_{\m\n}$. 

(3) Next, we obtain the gravitino equation by supersymmetric variation of
\eq{dual1}, and the $~\chi\-$field equation by supersymmetry variation of 
\eq{dual2}.

(4) Varying the gravitino equation under supersymmetry, we obtain the
Einstein equation \eq{b3}. In doing so, \eq{hs} given in the Appendix is 
useful in handling the $~H^2\-$terms. Note the occurrence of the trace
$~g_{\m\n}F^2\-$term in \eq{b3}, which is absent in the case of $~\nt=1$,
due to the use of dilaton equation of motion. Since, here we have multi
dilatons, the trace term can no longer be absorbed into the the dilaton
equation of motion.

(5) Varying the $~\chi\-$field equation \eq{f2}, we obtain the field
equation \eq{b4} for the generalized dilatons $~\varphi^{\un\a}$~ 
\cite{sagnotti}.

(6) Next, we obtain the hyperino field equation \eq{f3}, by the
requirement of the closure of two supersymmetries acting on the
hyperino $~\psi^a$.  Varying this equation under supersymmetry, we obtain
the field equation \eq{b5} for the hyperscalars $~\phi^\a$.

(7) Finally the gaugino field equation \eq{f4} is also obtained by the
closure of two supersymmetries on the gaugino $~\l$. The variation of
this equation under supersymmetry in turn yields the Yang-Mills
field equation \eq{b6}.

(8) Having determined the ungauged matter couplings and supersymmetry
transformation rules, we now turn on the gauge coupling constants
$~g,~g'$, thereby gauging the group $~Sp(\nh)\times Sp(1)$. To do this, we
follow the following steps:

(a) We gauge covariantize the relevant derivatives in the supersymmetry
transformation rules, as well as the field equations, according to the
rules described in section 2.

(b) Next, we find that the closure calculation will require the
introduction of only one new term to the transformation rules, namely
the $~C^{A B}\-$dependent term in the gaugino transformation rule \eq{t7}. 
This can be seen by examining the closure of supersymmetry on the gravitino, 
and by varying the new gauge coupling dependent terms in $~D_\m \e$.
Furthermore, we learn that we need to add the $~C^{A B}\-$dependent term in
the gravitino field equation.

(c) Having determined the fact that the gaugino transformation rule is
modified by the $~C^{A B}\-$dependent term that is proportional to the
gauge coupling constant, we then examine systematically the effect of
this new variation in all the closure calculations on the {\it fermions}.
Thus we determine all the new, gauge coupling constant dependent
modifications of the fermionic field equations, as given in \cite{ns}.

(d) Finally, we vary the new terms in the fermionic equations of motion
under the full transformation rules (old and new), as well as the old
terms in the fermionic equations of motion under the new, gauge coupling
constant dependent gaugino transformation rule, thereby obtaining all
the modifications, up to the fermionic bilinear terms in the bosonic
field equations of motion.

An important observation to be made here is that the gauging of the
$~SO(\nt,1)$~ or any of its subgroups does not work, because it is not
known how, and it may as well be impossible, to write down a gauge
covariant field strength for antisymmetric tensor fields.

\bigskip\bigskip
\centerline{\bf 4.~~Invariant Lagrangian for the Case of ~$\nt=1$}
\bigskip

When the number of self-dual tensor multiplets differs from that of
anti-self-dual tensor multiplets, the system lacks invariant lagrangian,
because we can not write down the kinetic term for purely (anti)
self-dual third-rank tensor. However, we {\it do} have an invariant
lagrangian for the case of $~\nt=1$~ as in refs.~\cite{ns,romans}. Since
this particular case is also of another importance with the geometry
$~SO(1,1)$~ with no isotropy group, we give the details of the system.

First of all, the coset space is now reduced to a semi-simple group
$~SO(1,1)$, and the coset representatives $~L\du I i$~ and $~L_I$~ can be
parametrized as
\be
\pmatrix{ L_0 &  L\du 0{(1)} \cr
L_1 &  L\du1{(1)} \cr}
= \pmatrix{ \cosh\theta & \sinh\theta\cr
\sinh\theta & \cosh\theta \cr} ~~, ~~~~
\pmatrix{ L^0 &  L^1 \cr
L\du{(1)}0 &  L\du{(1)} 1 \cr}
= \pmatrix{\cosh\theta & -\sinh\theta \cr
- \sinh\theta & \cosh\theta \cr }
\ee
for a rescaled field $~\theta\equiv \varphi^{\un 1}/{\sqrt2}$.
Accordingly, we have
\be
 \eta_{11} = + 1 ~~, ~~~~  \eta_{00} = - 1~~, ~~~~ \eta_{10} = 0 ~~,
\ee
\be
V\du{\ul 1}{(1)}\partial_\m\varphi^{\ul 1} 
= L^I \partial_\m L\du I{(1)} = \partial_\m\theta~~.
\ee
Following ref.~\cite{romans}, we define
\bea
&& a_{\m\n}\equiv\half \left( B_{\m\n}{}^0
- B_{\m\n}{}^1 \right) ~~,
~~~~ b_{\m\n} \equiv \half \left( B_{\m\n}{}^0 + B_{\m\n}{}^1 \right)~~, \nn\\
&& B_{\m\n}{}^0 = b_{\m\n} + a_{\m\n} ~~, ~~~~B_{\m\n}{}^1 =
b_{\m\n}-a_{\m\n}~~. 
\eea

We define the field strengths of $~a_{\m\n}$~ and $~b_{\m\n}$~ as
\bea
&& f_{\m\n\r}\equiv 3\partial_{\[\m} a_{\n\r\]}+3\tildevz \trz \left(
F_{\[\m\n} A_{\n\]} - \fracm 2 3 
A_{\[\m} A_\n{} A_{\r\]} \right) ~~, \nn\\
&& g_{\m\n\r} \equiv 3 \partial_{\[\m} b_{\n\r\]} + 3 v^z \trz \left(
F_{\[\m\n} A_{\n\]} - \fracm 2 3  A_{\[\m}
A_\n A_{\r\]} \right)\ , \la{deffg}  
\eea
where two {\it constants} $~\vz$~ and $~\tildevz$~ are defined by
\be
\vz \equiv \half\left( C^{0 z} + C^{1 z}\right)\ , ~~~~
\tildevz \equiv \half\left( C^{0 z} - C^{1 z}\right)\ . \la{vvt}
\ee
Accordingly we have  $C^{(1)z} = \vz e^\theta - \tildevz e^{-\theta}$, and
\be
C^z = \vz e^\theta + \tildevz e^{-\theta}\ . \la{cz}
\ee
After some manipulations, we get 
\bea
&& H_{\m\n\r}\equiv H\du{\m\n\r} I L_I = e^{-\theta} f_{\m\n\r}
+ e^{+\theta} g_{\m\n\r}
= 2e^{+\theta} g_{\m\n\r}^- ~~,~~~~ \nn\\
&& H\du{\m\n\r}{(1)}\equiv H\du{\m\n\r} I L_I{}^{(1)} 
= e^{+\theta} g_{\m\n\r} - e^{-\theta} f_{\m\n\r}
= 2e^{+\theta} g_{\m\n\r}^+~~. 
\eea
Due to the (anti)self-dualities of $~B_{\m\n}{}^I$, we have
$~f_{\m\n\r} = - e^{2\theta} \Tilde g_{\m\n\r}, ~
\Tilde f_{\m\n\r} = - e^{2\theta} g_{\m\n\r}$, where $~\Tilde g^{m n r}\equiv
(1/6) \e^{m n r s t u}g_{s t u}$~ and {\it idem} for $~\Tilde f^{m n r}$.

We can obtain the field equations of this system from our general
case, substituting above relations. The gravitino, dilatino,
gaugino, and hyperino field equations thus obtained are
\bea
&& \g^{\m\n\r} D_\n \psi\du\r A + \fracm1{12} e^\theta \g_{\[\m} \g^{\r\s\t} 
\g_{\n\]} \psi^{\n A} g_{\r\s\t} - \half \g^\n\g_\m \chi^A \partial_\n\theta 
-2 V\du\a{a A} \g^\n\g_\m\psi_a D_\n \phi^\a \nn\\
&& ~~ + \fracm12 C^z \g^{\r\s} \g_\m \trz\left(\l^A F_{\r\s}\right)
- \fracm1{12} e^\theta \g^{\r\s\t} \g_\m \chi^A g_{\r\s\t} - \trz\left(C^{A B} 
\g^\m \l_B \right) = 0 {~~,~~~} 
\la{gravitinoeq} \\[0.2in]
&& \g^\m \left( D_\m \chi -\half \g^\n\psi_\m \partial_\n\theta 
+ \fracm1{12} e^\theta \g^{\r\s\t} \psi_\m g_{\r\s\t}\right)  
- C_z^{-1} \left( \vz e^\theta - \tildevz e^{-\theta} \right) 
\trz\left( C^{A B} \l_B \right)  
\nn\\
&& ~~~~~ ~~~~~ - \fracm1{12}e^\theta \g^{\r\s\t} \chi\,
g_{\r\s\t} - \fracm12 (\vz e^\theta - \tildevz e^{-\theta}) \trz\left(
\g^{\r\s} \l F_{\r\s} \right) = 0 ~~,\\[0.2in]
&& C^z \g^\m \left( D_\m \l_A
+ \fracm14 \g^{\r\s} \psi_{\m A} F_{\r\s}  \right) 
+ \fracm14 \g^{\r\s} \chi_A (\vz e^\theta - \tildevz e^{-\theta} ) 
F_{\r\s}  \nn\\
&& ~~~~~+ \fracm12 (\vz e^\theta - \tildevz e^{-\theta})\g^\m\l_A\partial_\m
\theta + \fracm1{12} (\vz e^\theta -\tildevz e^{-\theta} ) e^\theta \g_{\r\s\t}
\l_A g_{\r\s\t} \nn\\
&& ~~~~~ +\half  C_{A B} \g^\m \psi\du\m B - \half C_z^{-1} 
\left( \vz e^\theta - \tildevz e^{-\theta} \right) C_{A B} \chi^B  
- 2\psi^a V\du{a A}\a \xi_\a = 0 ~~,\\[0.2in]
&& \g^\m \left( D_\m\psi^a - V\du\a{a A} \g^\n \psi_{\m A} D_\n \phi^\a \right) 
+ \fracm1{12} \g^{\r\s\t} \psi^a e^\theta g_{\r\s\t} 
- 2\l_A \xi^\a V\du\a {a A} = 0~~,
\eea
where $~\chi\equiv \chi^{(1)}$, and we recall that $~C^z
= \vz e^\theta + \tildevz e^{-\theta}$.

In a similar fashion, we can get all the bosonic field equations as
\bea
&& \fracm14 \left( R_{\m\n} - \half g_{\m\n} R \right)
- \fracm1{12} e^{2\theta} \left( 3g_{\m\r\s} g_{\n\r\s} - \half g_{\m\n}
g_{\r\s\t}^2  \right) - \fracm14 (\partial_\m\theta) (\partial_\n\theta)
+ \fracm18 g_{\m\n} (\partial_\r\theta)^2    \nn\\
&& 
~~~~~ ~~~ - \, g_{\a\b} (\partial_\m \phi^\a)(\partial_\n \phi^\b) + \half
g_{\m\n} g_{\g\d} g^{\r\s} (\partial_\r\phi^\g) (\partial_\s\phi^\d) 
+ \fracm 1 8 g_{\m\n} C_z^{-1} \trz \left(C^{A B} C_{A B} \right) \nn\\
&&
~~~~~ ~~~ - \fracm14 C^z \trz\left[\,
2F\du\m\r F_{\n\r} - \half g_{\m\n} (F_{\r\s})^2 \, \right] = 0 
{~~,~~}\nn\\[0.2in] 
&&
\half e^{-1}\partial_\m \left(e g^{\m\n}\partial_\n \theta \right)
- \fracm14 (\vz e^\theta - \tildevz e^{-\theta} )
F_{\m\n}^2 - \fracm16 e^{2\theta} g_{\r\s\t} ^2 \nn\\
&& ~~~~~ ~~~~~ ~~~~~ ~~~~~ ~~~~~ ~~  - \fracm 14 C_z^{-2} \left(\vz e^\theta 
- \tildevz e^{-\theta} \right) \trz\left(C^{A B} C_{A B}\right)= 0 ~~,\\[0.2in]
&& 
\fracm12 D_\n \left( \,e C^z F^{\m\n}\,\right) + \fracm12 e \left( \vz
e^{2\theta} g^{\m\r\s} F_{\r\s} - \tildevz \Tilde g^{\m\r\s} F_{\r\s}
\right) - e \xi_\a D_\m \phi^\a = 0 {~,~~~}\\[0.2in]
&& 
e^{-1} \partial_\m \left( e g^{\m\n} \partial_\n \phi^\a \right) +
g^{\m\n} \G_{\b\g}{}^\a (\partial_\m\phi^\b)(\partial_\n\phi^\g) 
- C_z^{-1} J\ud\a{\b A B} \trz \left( C^{A B} \xi^\b\right) = 0 ~~. 
\la{hyperoneq}
\eea
Now our antisymmetric tensor field equation is to be of the second order as a
combination of the self-dual and anti-self-dual parts of $~g_{\m\n\r}$:
To be more specific, we add the (anti)self-duality conditions
$~H_{\m\n\r}^-{}^I L\du I{(1)} + \cdots = 0$~ and $~H_{\m\n\r}^+{}^I
L_I + \cdots = 0 $~ in \eq{dual1} and {\eq{dual2}, to get
$~e^{2\theta} g_{\m\n\r} = - \Tilde f_{\m\n\r} + \cdots$. We next take
its divergence to get
\be 
\half D_\m \left(e e^{2\theta} g^{\m\n\r}\right)
= - \fracm18 \tildevz \e^{\n\r\m\s\t\o} \trz ( F_{\m\s} F_{\t\o} )  ~~.
\ee

Similarly the supersymmetry transformation rule is also derived as 
\bea
&& \d e\du\m m = + (\bar\e\g^m \psi_\m) ~~, \nn\\ 
&& \d \psi_\m = + D_\m(\Hat\o,A_{\un\a}) \e 
+ \fracm1{24} e^\theta \g^{\r\s\t} \g_\m \e
g_{\r\s\t} - (\d\phi^\a)(A_\a \psi_\m) ~~ \nn\\ 
&& ~~~~~ ~~~ - \fracm1{16} \g_\m\chi (\Bar\e \chi) - \fracm3{16} \g_\n\chi 
(\Bar\e \g_{\m\n} \chi) + \fracm1{32} \g_{\m\n\r} \chi
(\Bar\e\g^{\n\r}\chi) 
+ \fracm1{16} \g^{\n\r} \e (\Bar\psi{}^a \g_{\m\n\r} \psi_a) \nn\\
&& ~~~~~ ~~~ - \fracm1{16} C_z
\trz \left [ \, 18 \l (\Bar\e \g_\m\l) - 2 \g_{\m\n}\l (\Bar\e\g^\n\l )  
+ \g_{\r\s} \l (\Bar\e\g\du\m{\r\s}\l ) \, \right] ~~, \nn\\[0.15in]
&& \d b_{\m\n} = +2 \vz \trz \left( A_{\[\m} \d A_{\n\]} \right)   
- e^{-\theta} (\Bar\e \g_{\[\m} \psi_{\n\]} ) 
+ \half e^{-\theta} (\bar\e \g_{\m\n} \chi ) ~~, ~~~~ 
\d\theta = + ( \Bar\e\chi ) ~~, \nn\\ 
&& \d \chi = + \half \g^\m \e\partial_\m \theta - (\d\phi^\a)(A_\a\chi) 
- \fracm1{12} e^\theta \g^{\m\n\r} \e g_{\m\n\r} 
+ \fracm12 \left( \vz e^\theta-\tildevz e^{-\theta} \right) 
\trz \left[ \, \g^\m \l (\Bar\e \g_\m \l ) \, \right] ~~ , \nn\\[0.15in] 
&& \d A_\m  = +  ( \Bar\e \g_\m \l )~~, \nn\\ 
&& \d \l_A = - \fracm1 4 \g^{\m\n} \e_A F_{\m\n} - ( \d\phi^\a) (A_\a
\l_A ) - C_z^{-1} \left(\vz e^\theta - 
\tildevz e^{-\theta} \right) \left( \Bar\chi_{(A} \l_{B)} \right)\e^B 
- \half C_z^{-1} C_{A B} \e^B ~~, \nn\\[0.15in]  
&& \d \phi^\a = + V\du{a A}\a (\Bar\e{}^A \psi^a) ~~, \nn\\ 
&& \d\psi^a = + V\du\a{a A} \g^\m \e_A \Hat D_\m\phi^\a - ( \d\phi^\a) (A_\a
\psi)^a ~~. 
\la{n1trsf}
\eea

Note that in the absence of gauging, the function $C^{A B}$ vanishes and
singular behaviour in the couplings arises in the energy-momentum tensor
for the Yang-Mills field in \eq{b1} and in the Yang-Mills equation
\eq{b3}. Furthermore, the $\l^2\epsilon$ terms in the supersymmetry
variation of the gravitino vanish and the $\chi\l\epsilon$ terms in the
supersymmetry variation of the gaugino diverge, at the critical point
where $C^z = \vz e^\theta + \tildevz e^{-\theta}$ vanishes. When the
gauging of $Sp(\nh)\times Sp(1)$ is switched on, the divergent
$C_z^{-1}$ factors arise in $\chi$, $\l$, Einstein, dilaton and
hypermatter field equations, and the supersymmetry variation of the
gaugino picks up another singular contribution.

When $~\tildevz = 0$, this result agrees with ref.~\cite{romans} as far
as the supergravity and Yang-Mills multiplets are concerned, and also
with ref.~\cite{ns} with hypermultiplets. When $~\vz=0$, this result
coincides with the system in ref.~\cite{ng6d}. In the important cases of
$~\vz=0$~ and $~\tildevz =0$, this is reduced to the usual exponential
factors. It is worthwhile to mention that the $~\chi\l\-$terms in the
~$\l\-$transformation rules have apparently different coefficients
compared with \cite{ns}, {\it via} \eq{lc}. This is attributed to the
fact that our gaugino ~$\l\-$field is rescaled by an exponential
function of the dilaton ~$\theta$. 

As will be discussed in the next section, since the conservation of the
Yang-Mills current is satisfied only for $\eta\low{I J}C^{I z} C^{J
z'}=0$, the invariant lagrangian exists only for the two cases
$~\vz=0$~ or $~\tildevz=0$~ for our $~\nt=1$~ system. If we formally try
to integrate the field equations for other cases of $~\vz\Tilde v{}^{z'}
\neq0$, we encounter gauge non-invariant terms in the lagrangian, that
invalidate supersymmetry. This is because commutators of two
supersymmetries will result in a gauge transformation. Another explicit
way to see this is to take the variation of the $~b\wedge F\wedge F$~
term in the lagrangian under supersymmetry. This produces gauge
non-invariant terms proportional to $~\vz\Tilde v{}^{z'}$ which can not
be cancelled. Note also that this feature for the case $~\vz\Tilde
v{}^{z'} \neq 0$~ at the classical level does {\it not} necessarily mean
the system is inconsistent, as will be discussed in the next section.

In the case of $~\tildevz=0$, the invariant lagrangian by integrating the field
equations is:
\bea
e^{-1} \Lag_{n=1}^{\Tilde v=0}&=& +\fracm14 R(\o) - \fracm1{12} e^{2\theta}
g_{\r\s\t}^2 - \fracm14 (\partial_\m\theta)^2 -  \half\left( \Bar\psi_\m
\g^{\m\n\r}D_\n \psi_\r \right) - \half \left(\Bar\chi \g^\m D_\m \chi\right)
\nn\\ 
&& - \fracm14 \vz e^\theta \trz (F_{\m\n})^2 - \vz e^\theta
\trz \left(\Bar\l \g^\m D_\m \l \right) \nn\\
&& - g\low{\a\b} g^{\m\n} (\partial_\m\phi^\a) (\partial_\n\phi^\b)
- \left( \Bar\psi{}^a \g^\m D_\m \psi_a \right) 
+ \fracm12 \vz e^\theta \trz (\Bar\chi \g^{\m\n} \l F_{\m\n}) \nn\\ 
&& + \half \left(\Bar\psi_\m \g^\n\g^\m \chi \right) \partial_\n \theta 
+ 2 \left(\Bar\psi_{\m A} \g^\n\g^\m \psi_a\right) V\du\a{a A}
\partial_\n\phi^\a \nn\\ 
&& - \fracm12 \vz e^\theta 
\trz \left(\Bar\psi_\m \g^{\r\s} \g^\m \l F_{\r\s} \right) 
\nn\\  
&& - \fracm1{24} e^\theta g_{\m\n\r} \bigg[ \left(\Bar\psi{}^\l\g_{\[\l}
\g^{\m\n\r} \g_{\t\]} \psi^\t\right) - 2 \left( \bar\psi_\l \g^{\m\n\r} \g^\l
\chi \right) \nn\\ 
&& ~~~~~ ~~~~~ ~~~~~ ~~~ + 2 \left(\Bar\psi{}^a \g^{\m\n\r}\psi_a \right) 
- \left(\Bar\chi \g^{\m\n\r}\chi \right) 
+ 2 \vz e^\theta \trz \left( \Bar\l \g^{\m\n\r} \l \right) \bigg] \nn\\ 
&& - \fracm14 v_z^{-1} e^{-\theta} \trz\left(C^{A B} C_{A B} \right) \nn\\
&& - 4 \trz\left( \Bar\psi_a\l_A V\du\a{a A} \xi^\a \right) 
+ \half \trz\left( \Bar\psi\du\m A \g^\m \l^B C_{A B}\right) 
- \trz\left( \Bar\chi_A\l_B C^{A B} \right) ~~ . 
\la{tildev0lag}
\eea
This lagrangian with the transformation 
rule \eq{n1trsf} with $~\tildevz =0$~ coincides with the system in
refs.~\cite{romans,ns} with the Chern-Simon modification in the
$~b_{\m\n}\-$transformation and its field strength.  In 10D, this corresponds
to the system in ref.~\cite{cm}.  

In the case $~\vz=0$, the invariant lagrangian is 
\bea
e^{-1} \Lag_{n=1}^{v=0} &=& + \fracm14 R(\o) - \fracm1{12} e^{2\theta}
g_{\r\s\t}^2 - \fracm14 (\partial_\m\theta)^2 -  \half\left( \Bar\psi_\m
\g^{\m\n\r}D_\n \psi_\r \right) - \half \left(\Bar\chi \g^\m D_\m \chi\right)
\nn\\ 
&& - \fracm14 \tildevz e^{-\theta}  \trz (F_{\m\n}) ^2 - \tildevz e^{-\theta} 
\trz \left(\Bar\l \g^\m D_\m \l \right) \nn\\
&& - g\low{\a\b} g^{\m\n} (\partial_\m\phi^\a) (\partial_\n\phi^\b)
- \left( \Bar\psi{}^a \g^\m D_\m \psi_a \right) 
- \fracm12 \tildevz e^{-\theta} \trz (\Bar\chi \g^{\m\n} \l 
F_{\m\n}) \nn\\ 
&&  + \half \left(\Bar\psi_\m \g^\n\g^\m \chi \right) \partial_\n \theta 
+ 2 \left(\Bar\psi_{\m A} \g^\n\g^\m \psi_a\right) V\du\a{a A}
\partial_\n\phi^\a \nn\\ 
&& - \fracm12 \tildevz e^{-\theta} 
\trz\left( \Bar\psi_\m \g^{\r\s} \g^\m \l F_{\r\s} \right)
+ \fracm18 \tildevz e^{-1} \e^{\m\n\r\s\t\o} b_{\m\n} 
\trz ( F_{\r\s} F_{\t\o} ) 
\nn\\  
&& - \fracm1{24} e^\theta g_{\m\n\r} \bigg[ \left(\Bar\psi{}^\l\g_{\[\l}
\g^{\m\n\r} \g_{\t\]} \psi^\t\right) - 2 \left( \bar\psi_\l \g^{\m\n\r} \g^\l
\chi \right) \nn\\ 
&& ~~~~~ ~~~~~ ~~~~~ ~~~ + 2 \left(\Bar\psi{}^a \g^{\m\n\r}\psi_a \right) 
- \left(\Bar\chi \g^{\m\n\r}\chi \right) - 2 \tildevz e^{-\theta}   
\trz \left( \Bar\l \g^{\m\n\r} \l \right) \bigg] \nn\\
&& - \fracm14 {\Tilde v}_z^{-1} e^{-\theta} \trz\left(C^{A B} C_{A B} \right)
\nn\\
&& - 4 \trz\left( \Bar\psi_a\l_A V\du\a{a A} \xi^\a \right) 
+ \half \trz\left( \Bar\psi\du\m A \g^\m \l^B C_{A B}\right)  
+ \trz \left( \Bar\chi_A\l_B C^{A B} \right) ~~ . 
\la{v0lag}
\eea 
This lagrangian and the transformation  
with $~\vz=0$~ correspond to the system in refs.~\cite{ng6d,romans} with 
the explicit $~b\wedge F\wedge F\-$term in the lagrangian with no
modification in the $~b_{\m\n}\-$transformation rule or its field strength.  
In 10D, this corresponds to the dual formulation \cite{ch}.  

\bigskip\bigskip\bigskip

\centerline{\bf 5.~~Discussion}
\bigskip

In this paper we have constructed the field equations of the combined
system of the $~N=1$~ supergravity multiplet, $~\nt$~ copies of
anti-self-dual tensor multiplets with anti-self-dual tensor multiplet,
forming the coset space $~SO(\nt,1)/SO(\nt)$, Yang-Mills multiplets, and
hypermultiplets forming the coset $~Sp(\nh,1)/Sp(\nh)\otimes Sp(1)$.
Furthermore we have gauged the group $~Sp(\nh)\times Sp(1)$. These are the
most general couplings of the six dimensional supergravity plus matter
system to date. The resulting system exhibits some interesting features
which we now comment on.

We have already commented on the singular behaviour of the couplings at
a special point in moduli space, and the occurrence of new divergent
couplings at the same point which proportional to the gauged
$Sp(\nh)\times Sp(1)$ coupling constants. Another important feature,
which was observed in \cite{ferraraetal}, and which continues to hold in
the more general system presented here, is the anomalous behaviour of
the gauge couplings. Namely, writing the Yang-Mills equation \eq{b6} as 
\be
{\cal D}_\n \left(e C^z F^{\m\n}\right) = e J^\m\ , \la{dj}
\ee
we find that
\be
{\cal D}_\m (e J^\m) = \fracm 1 {16} \e^{\m\n\r\s\t\l} \eta\low{I J}
C^{I z} C^{J z'} F_{\m\n} \tr_{z'} \left( F_{\r\s} F_{\t\l} \right)\ . 
\la{ae}  
\ee
Perhaps not too surprisingly, the hypermatter contributions to this
anomaly equation have all canceled.  Setting 
\be
\eta\low{I J}C^{I z} C^{J z'}=0 \ , \la{c2}
\ee
eliminates the anomalous divergence. This corresponds to the familiar
$~\nt=1$~ case \cite{ns} for which a covariant action can be written
down. Interestingly, there are two possibilities that have been treated
simultaneously here. In notation of section 4, these correspond to the
cases of $~\vz=0~$ or $~\tildevz=0$. As discussed in section 4, the case
of $~\tildevz=0$~ is the familiar one constructed by the
authors a long tome ago \cite{ns}, and its ten dimensional analog is well-known
\cite{peter,cm}. The case of $~\vz =0~$ corresponds to the dual
formulation, constructed also long ago \cite{ng6d}, and its ten
dimensional analog is the dual formulation of Chamseddine \cite{ch}.

It should be emphasized that neither the elimination of the anomalous
divergence \eq{ae} ensure anomaly freedom, nor does its nonvanishing
mean that the theory is anomalous. The property of anomaly freedom
solely depends on the choice of matter multiplets, and as is well-known, there
are many available anomaly free sets of such multiplets, some of which
will be discussed below. 

Considering the case of $~\nt > 1$, eq.~\eq{dj} represents the Bose
non-symmetric covariant gauge anomaly, as observed in
\cite{ferraraetal}. Its Bose symmetric covariant version, as well as the
associated local supersymmetry anomaly can be determined by
considerations of Wess-Zumino anomaly consistency conditions
\cite{ferraraetal}. Although a covariant action can not be written down
for $~\nt> 1$, both the gauge as well as supersymmetry anomalies can
nonetheless be associated with the gauge and supersymmetry variation,
respectively, of the Green-Schwarz type term \cite{ferraraetal}
\be
\Lag_{GS} = -\fracm 1 8  \e^{\m\n\r\s\t\l} \left( \eta\low{I J}
B_{\m\n}{}^I C^{J z} \right) \tr_{z} \left( F_{\r\s} F_{\t\l} \right)
-\half \e^{\m\n\r\s\t\l}\eta\low{I J} C^{I z} C^{J z'} \omega_{\m\n\r}^z
\omega_{\s\t\l}^{z'}\ , \la{gs}
\ee
where $~\omega_{\m\n\r}^z$~ is the Yang-Mills Chern-Simons form. 

The results of \cite{ferraraetal} and our generalization which includes
the hypermatter reveal therefore, a surprising situation in which an
anomalous system of supermultiplets {\it can} be supersymmetrized, with
the caveat that the integrability conditions for the equations of motion
reflect the anomalies. The fact that this is possible at all may be due
to the manifest gauge covariant nature of the field equations and
supersymmetry transformation rules.  Attempting to supersymmetrize 
a gauge non-invariant action, on the other hand, 
would run immediately into trouble with supersymmetrization.

The anomalies of the full system discussed here are to be cancelled by
the quantum one-loop effects, so that the total effective action is
gauge invariant and supersymmetric, provided that the right set of
matter multiplets are included. The anomalies may cancel precisely, or
Green-Schwarz cancellation mechanism may have to be employed for the
cancellation \cite{gs}. The same mechanism works in the dual formulation
as was shown in \cite{gn,ss,ng6d}. In the case of $~\nt=1$, a generalized
version of the Green-Schwarz mechanism was found by Sagnotti
\cite{sagnotti} which involves the use of multi-tensor fields
simultaneously. We conclude by mentioning some of the anomaly free
matter contents for the $~\nt=1$~ and $~\nt>1$~ cases.

For ~$\nt=1~$ without gauging, a large number of anomaly free matter
contents can be obtained by compactifying the well-known anomaly free
ten dimensional supergravity-Yang-Mills systems that arise in string
theory \cite{gsw}. All of these models have a gauge group of rank ~$\le
20$, and they arise from a perturbative treatment of string
compactification. Witten \cite{witten} has discovered a new mechanism by
which a nonperturbative symmetry enhancement occurs, and a new class of
anomaly-free models, not realized in perturbative string theory, emerge
in 6D. These can have rank greater than 20. Schwarz
\cite{sc} has constructed new anomaly-free models in 6D, some of which may
potentially arise in a similar nonperturbative scheme.

As for the gauged case with ~$\nt=1$, an anomaly free model was found in
\cite{rsss}, where the gauge group is ~$E_6\times E_7\times U(1)$. The
~$U(1)$~ factor is a subgroup of the automorphism group, and the
hyper-fermions belong to the ~$912$~ dimensional representation of ~$E_7$.
The origin of this model still remains mysterious, and it would be very
interesting to determine if it can be explained by a new kind of
nonperturbative mechanism in M-theory.

In general, the necessary but not sufficient condition for the anomaly
cancellation is \cite{rsss}
\be
\nh - \nv + 29\, \nt = 273\ .
\ee
As mentioned in the introduction, an example of an anomaly free model
with ~$\nt =9~$ has been found in \cite{sen} by considering a suitable
M-theory compactification, and it has ~$n{\low V}=8,~\nh=20$. The matter
couplings of six dimensional supergravity constructed here provides the
field theoretic description of this model.

Finally, we mention an example of an anomaly free matter content with 
~$\nt>1~$ found sometime ago in \cite{bkss}. It has: 
\be
\nt=17\ , \quad \nh=28\ , \quad \nv =248 \ .
\ee
The vector fields fit into the adjoint representation of ~$E_8$, and we
can take the hyperscalar manifold to be ~$E_8/(E_7\times Sp(1))$, in
which case the hyperfermions transform in ~$56~$ dimensional
representation of ~$E_7$. As far as we know, this model, which has a
rather simple field content, has not found an M-theory explanation so
far, and it would be interesting to see if there is one.

\bigskip\bigskip

\centerline{\bf Acknowledgements}
\bigskip

The authors are grateful to M.~Duff, A.~Sagnotti and E.~Witten for
stimulating discussions. H.N. would like to thank the Center for
Theoretical Physics at Texas A\& M University for hospitality.

\vfill\eject

\centerline{\bf Appendix:~~Notations, Conventions and Lemmas}

\bigskip

Our metric is $~(\eta_{m n}) = \hbox{diag.}~(-,+,+,+,+,+)$, while the
Clifford algebra is generated by $~ \{ \g_m, \g_n\} = 2\eta_{m n}$. Note
that this signature differs from the one in \cite{ns}. The definition of
the Ricci tensor is the same as in \cite{ns}, however, namely:
$R_{\m}{}^a= R_{\m\n}{}^{ab}\, e_b{}^\n{}$. We define
$~\g_7 \equiv \g_{(0)\, (1)\, \cdots\, (5)},~ \e^{0 1 2 3 4 5} = + 1$, 
such that $~(\g_7)^2=+1$.  More generally we have
\be 
\g^{r_1\cdots r_n} = \fracm{(-1)^{\[ n/2 \]}}
{(6-n)!} ~ \e^{r_1\cdots r_n s_1\cdots s_{6-n}} \g_{s_1\cdots s_{6-n}}
\g\low 7 ~~.
\ee
The basic gamma-matrix relations such as $~\g_m \g^{r s t} \g^m=0$~
stays the same as in ref.~\cite{ns}, as well as the conventions
for the $~Sp(1)$~ indices, {\it e.g.},
\be 
\chi^A{}_i = \e^{A B} \chi\low{B i} ~~, ~~~~
\chi\low{A i} = \chi^B{}_i \e\low{B A}~~,  ~~~~
\left(\e\low{A B}\right) = \left(\e^{A B}\right) = \pmatrix{ 0 & 1 \cr -1 
& 0 \cr } ~~,  \ee
\be 
\chi^A{}_i = \e^{A B} {\Bar\chi}\low{B i}{}^T ~~, ~~~~ \Bar\chi\low{A i}
= (\chi^A{}_i)^\dagger \g_0~~,  
\ee
\be 
\left( \Bar\chi{}^A{}_i \g^{m_1\cdots m_n} \l^B\right) = (-1)^{n+1}
\left( \Bar\l {}^B \g^{m_n\cdots m_1} \chi^A{}_i \right) ~~.
\ee
For inner products of $~Sp(1)$~ (or $Sp(\nh)$) symplectic spinors
\cite{kt}, the contractions with $~\e\low{A B}$~ (or $~\e_{a b}$) are
always understood, {\it e.g.,} $~\left( \Bar\chi_i\g^{r s} \l\right) =
\left(\Bar\chi^A{}_i \g^{r s} \l_A\right)$~ as in \cite{ns}, {\it e.g.,}
\be 
\left(\Bar\chi\,\low i \g^{r_1\cdots r_n} \l\right) = (-1)^n
\left(\Bar\l \g^{r_1\cdots r_n} \chi\,\low i\right) ~~.
\ee
Exactly as in ref.~\cite{ns}, for given four symplectic Majorana-Weyl spinors
$~\psi_1, ~\cdots, ~\psi_4$, where the labels $~{\scst 1,~\cdots,~4}$~ denote
{\it all} the possible indices they may carry, including $~Sp(1)$,
$~Sp(\nh)$~ or $~SO(\nt)$~ indices, the Fierz arrangement formula is
\bea
\left( \Bar\psi{}_1 \psi_2 \right)\left( \Bar\psi{}_3
\psi_4 \right) 
 &=&- \fracm 1 8 (1+ c_2 c_4) \left[\, \left(
\Bar\psi {}_1 \psi_4 \right) \left( \Bar\psi_3 \psi_2\right) -
\half\left( \Bar\psi{}_1 \g^{rs} \psi_4 \right)
\left({\Bar\psi}{}_3 \g_{r s} \psi_2\right) \, \right] \nn\\
&&-\fracm 1 8 (1-c_2 c_4 ) \left[ \, \left( \Bar\psi{}_1 \g^r \psi_4\right)
\left(\Bar\psi{}_3 \g_r \psi_2\right) - \fracm 1{12} \left( \Bar\psi{}_1
\g^{r s t} \psi_4 \right) \left( \Bar\psi{}_3 \g_{r s t}
\psi_2\right) \, \right] \quad\quad 
\eea
where $~\g\low7 (\psi_2 , \psi_4) = (c_2 \psi_2,~ c_4 \psi_4)$.

One of the most frequently used relationships related to the
(anti)self-dual tensors is $~S^{\m\n\r} S_{\m\n\r}\equiv
A^{\m\n\r}A_{\m\n\r} \equiv 0$, where the third-rank tensors $~S$~ and
$~A$~ are respectively self-dual and anti-self-dual tensors: $~(1/6)
\e\du{m n r} {s t u} S_{s t u} = + S_{m n r}, ~(1/6) \e\du{m n r} {s t
u} A_{s t u} = - A_{m n r}$. For the tensor $~H_{\m\n\r}{}^I$~ we use
the symbols $~H_{\r\s\t}^+{}^I$~ (or $~H_{\r\s\t}^{-}{}^I$) to
distinguish their dual (or anti-self-dual) components. The important
duality properties of the gamma-matrices multiplied by fermions are
summarized as follows. For fermions with the positive chirality such as
$~\psi_\m{}^A,~\l^{A\, r}$~ or $~\e^A$, or for fermions with negative
chiralities such as $~\chi^{A \,i},~\psi^a$, we have
\be
 \fracm16 \e\du{m n r}{s t u} \left( \g_{s t u} \psi_\m\right) \equiv -
\left( \g_{m n r} \psi_\m \right)  ~~, ~~~~
\fracm16 \e\du{m n r}{s t u} \left( \g_{s t u} \psi^a \right) \equiv +
\left(\g_{m n r} \psi^a \right)  ~~,
\ee
In other words, the combination $~\g^{r s t} \psi_\m$~ behaves as an
anti-self-dual tensor, while $~\g^{r s t} \psi^a$~ behaves as a
self-dual tensor, as far as the indices $~{\scst \[r s t\] }$~ are
concerned. It also follows that $~\g^{\r\s\t}\l H_{\r\s\t}^-{}^I \equiv
0$~ or $~\g^{\r\s\t}\chi H_{\r\s\t}^+{}^I \equiv 0$.

In the remainder of this appendix, we shall list a number of lemmas that
are useful in the derivation of field equations and supersymmetry
transformation rules.

(1) For $~H^2\-$term computation in gravitational equation 
the following lemma is useful:
\be
H^+_{\[\m\n}{}^{\tau i} H^+_{\r\]\s\tau i} \equiv 0 ~~. \la{hs}
\ee
This can be verified by using the duality property of ~$H$, and simple
manipulations involving the Schouten identity, which in the present case
means that an antisymmetrization of seven world indices vanishes identically.

(2) For $~\chi H\-$terms in the derivations of $~\chi\-$field equation
out of anti-self-duality condition, the following lemma is useful:
\be
\left( \Bar\e\g\ud\s{\[\m|}\chi^i \right) H_{|\n\r\]\s}^-
\equiv - \fracm1 3 \left( \bar\e\chi^i\right) H_{\m\n\r}^- ~~.
\ee
This can be proven by the vanishing of $~\left( \Bar\e\g_{\m\n\r} \g^{\s\t\o}
\chi^i \right) H_{\s\t\o}^- = 0 $~ with the $~\g\-$algebra for
the l.h.s.

(3) In calculating the divergence of the Yang-Mills field equation, it
is useful to note that
\be
D_\m C^z= \partial_\m C^z= (\partial_\m \varphi^{\un\a}) V_{\un\a}{}^i 
C^z_i\ , 
\quad\quad\quad D_\m C^{z i} = (\partial_\m \varphi^{\un\a}) V_{\un\a}{}^i
C^z\ . 
\ee

(4) For the gravitino field equation out of self-duality condition, we use the
lemma
\be
\left( \Bar\e\g^\s \g_{\m\n\r} \g^{\t\omega} \chi_i \right)
H_{\s\t\omega}^+{}^i = - \fracm 1 3
\left(\Bar\e \g^{\s\t\omega} \g_{\m\n\r} \chi_i \right)
H_{\s\t\omega}^+{}^i - 16 \left(\Bar\e \chi_i \right) H_{\m\n\r}^+{}^i ~~,
\ee
confirmed by $~\g\-$algebra as well as the self-duality of $~H^+$.

(5) The $~\l^2$~ and ~$\chi^2\-$terms in the supersymmetry transformation of
the gravitino can be rearranged as
\bea
&& \d\psi_\m \big|_{\l^2} = - C^{I z}L_I \left(
\fracm 3 4 \e \low B \l\du\m { B A}
+ \fracm 1 4 \g\du\m\n \e\low B \l\du\n {B A}
+ \fracm 1{16} \g^{\s\t} \e^A \l_{\m\s\t} \right) ~~, \\
&&\d\psi_\m \big|_{\chi^2} = - \fracm3 8 \e\low B \chi\du\m{B A} + \fracm 18
\g\du\m\n \e\low B \chi\du\n{B A} ~~,
\eea
where $~\l\du\m{A B} = \tr_z \left( \Bar\l{}^A\g_\m
\l^B \right), ~ \l_{\r\s\t} \equiv \tr_z\left(\Bar \l\g_{\r\s\t} \l \right), ~
\chi\du\m{A B} \equiv \left(\Bar\chi {}^A \g_\m\chi{}^B\right) ,~
\chi_{\r\s\t} \equiv \left(\Bar\chi\g_{\r\s\t} \chi \right)$.

(6) The $~\l\chi\-$term in $~\d\l$~ \eq{t7} can be rewritten 
by using the identity
\be 
\left(\Bar\chi_{i(A}\l_{B)} \right) \e^B =
\fracm 1 4 \l_A \left(\Bar\e \chi_i\right) + \fracm 1 8  \g_{\r\s} \l_A
\left(\Bar\e\g^{\r\s} \chi_i \right) 
+ \fracm 1 2 \e_A \left(\Bar\l \chi_i\right)~~, \la{lc} 
\ee
obtained by Fierz rearrangement.

(7) In order to fix the $~\l^2\-$terms in the supersymmetry transformation 
of the gravitino, we arrange the $~\psi^a\l^2\-$terms in the commutator on
$~\psi^a$, which needs the lemma
\bea
&& 
\left( \Bar\e_2^A \g_{\m\n\r} \e_{1 B} \right)
\l\ud {\r B} A = - 4 \tr_z \left(\Bar\e_{\[ 1|} \g_\m \l \right)
\left(\Bar\l\g_\n \e_{|2\]}  \right) + \half \xi^\r \l_{\m\n\r} ~~,\nn\\
&& 
\tr_z \left(\Bar\e_{\[ 2|} \g^{m n}{}_\r \l\right)\left(\Bar\e_{|1\]} \g_\r\l
\right) = - \half \xi^\r
\l\du\r{m n} + 2 \tr_z\left( \Bar\e_{\[2|} \g^{\[ m} \l \right)
\left(\Bar\e_{|1\]} \g^{n\]} \l \right)  ~~, {~~~~~}
\eea
where $~\xi^\m \equiv \left( \Bar\e_2\g^\m \e_1\right)$.

(8) The following non-trivial lemmas are useful for the closure checks on
$~e\du\m m$~ or $~B_{\m\n}$:
\bea
&& 
\trz \left(\Bar\e_1 \g_{\r\s\[\m} \l\right) \left( \Bar\e_2
\g_{\n\]}{}^{\r\s} \l\right) - {\scst (1 \leftrightarrow 2)} \equiv 0 ~~,
~~~~ \left(\Bar\e_1 \g^{\m\n\r\s} \chi^{(i} \right) \left( \Bar\e_2 \g_{\r\s}
\chi^{j)} \right) - {\scst (1 \leftrightarrow 2)}\equiv 0 ~~, \nn\\
&&
\left(\Bar\e_2 \g^\r \e_1 \right) \trz \left( \Bar\l\g_{\m\n\r} \l
\right) \equiv \trz \left[ \, 2 \left( \Bar\e_2 \g_\m \l \right)\left(\Bar\e_1
\g_\n \l \right) - \left( \Bar\e_2 \g_{\m\n\r} \l \right) \left(
\Bar\e_1 \g^\r \l \right)  \, \right] 
- {\scst (1 \leftrightarrow 2)}~~. \la{eb}
\eea
These are easily confirmed by appropriate Fierz arrangements as well as the
duality properties we already know.

(9) It is useful to note the following relation for the closure check on $~\l$:
\bea
&&
\left(\Bar\e_1^{(A} \g_{\s\t\r} \e_2^{B)} \right) \g^{\m\n\s\t}
\l_B H_{\m\n\r}^- \equiv 0  ~~,\nn\\
&&
\left(\Bar\e_1^{(A} \g^{\m\r\s} \e_2^{B)} \right) \g\du\m\n \l_B
H_{\n\r\s}^- \equiv 0  ~~,
\eea
which can be proven by the relationship
$~ \bigg( \Bar\e{}_1^{(A} \g_{\r\s\t} \e_2^{B)} \bigg) \[ \g^{\r\s\t},
\g^{\m\n\o} \]_\pm \l H_{\m\n\o}^- \equiv 0$, {\it etc.}
\newline due to the
anti-self-duality of the combination $~\left( \Bar\e_1^A \g_{\r\s\t}\e_2^B
\right)~$ as well as the anti-self-duality of $~H^{-\,I}$.

(10) In the arrangement of $~\l\chi^2\-$terms in the commutator on $~\l$, the
following lemma for super-variation is useful:
\be
\d \left(C_z^{-1}C^{i z} \right) =
C_z^{-2} \left(\Bar\e\chi^j  \right) \left( \d^{i j}
C_z^2 - C^{i z} C^{j z}\, \right) ~~.
\ee

\pagebreak

\ed